\documentclass[10pt,a4paper,english,aps,prd,superscriptaddress,preprintnumbers,showpacs,notitlepage,eqsecnum,nofootinbib]{revtex4-1}
\usepackage[latin9]{inputenc}
\usepackage{xcolor}
\usepackage{amsmath}
\usepackage{esint}

\makeatletter

\usepackage{hyperref}
\usepackage[english]{babel}
\usepackage{amssymb}
\usepackage{graphicx}
\usepackage{xcolor}
\usepackage{eufrak}
\usepackage{caption}
\usepackage{subcaption}
\def\sqr#1#2{{\vcenter{\vbox{\hrule height.#2pt\hbox{\vrule
width.#2pt height#1pt \kern#1pt\vrule width.#2pt}\hrule height.#2pt}}}}

\@addtoreset{equation}{section}

\makeatother

\usepackage{babel}
\begin{document}
\vglue 1cm

\title{Thermodynamics sheds light on black hole dynamics}

\author{Marcela C\'ardenas}
\affiliation{APC, Université Paris Diderot,~\\
 CNRS, CEA, Observatoire de Paris, Sorbonne Paris Cité~\\
 10, rue Alice Domon et Léonie Duquet, F-75205 Paris CEDEX 13, France.}
 \affiliation{Universidad Católica del Maule,\\
Av. San Miguel 3605, Talca, Chile.}
\author{Félix-Louis Julié}
\affiliation{APC, Université Paris Diderot,~\\
 CNRS, CEA, Observatoire de Paris, Sorbonne Paris Cité~\\
 10, rue Alice Domon et Léonie Duquet, F-75205 Paris CEDEX 13, France.}
\author{Nathalie Deruelle}
\affiliation{APC, Université Paris Diderot,~\\
 CNRS, CEA, Observatoire de Paris, Sorbonne Paris Cité~\\
 10, rue Alice Domon et Léonie Duquet, F-75205 Paris CEDEX 13, France.}

\date{December 7th, 2017}
\begin{abstract}
We propose to unify two a priori distinct aspects of black hole physics : their thermodynamics,
 and their effective dynamics when they are ``skeletonized" as point particles (a useful procedure when tackling, for example, their motion in a coalescing binary system). For that purpose, the Einstein-Maxwell-dilaton (EMD) theory, which contains simple examples
of asympotically flat, hairy black hole solutions, will serve as a laboratory. We will find that, when reducing a black hole to a point particle endowed with its specific, scalar-field-sensitive, effective mass, one in fact describes a black hole satisfying the first law of thermodynamics, such that its global charges, and hence its entropy, remain constant. This shows that the integration constant entering the scalar-field dependent mass is its entropy. \end{abstract}

\maketitle
\vskip 0.5cm
\section{Introduction}

Thermodynamics has proven to be a powerful tool to give a physical interpretation of the integration constants characterizing a black hole spacetime, in introducing its extensive parameters (global charges and entropy) and its intensive ones defined on the horizon (temperature, electric potential, etc.). The first law then tells us how a black hole readjusts its equilibrium configuration when interacting with its environment.\\

 On the other hand, the dynamics of interacting (non rotating) compact objects often relies on their ``skeletonization", that is, on reducing them to effective point particles endowed with a constant ``mass" parameter. In the case of a general relativistic Schwarzschild black hole, the interpretation of this parameter is straightforward as it cannot be anything else than its Schwarzschild mass.\\
  
   Consider now Einstein-Maxwell-dilaton theories, which consist in supplementing general relativity with a scalar field and a (non-minimally coupled) vector field. Such theories allow for the existence of hairy black hole solutions. Their reduction to point particles was recently performed in \cite{Julie:2017rpw} and involves a scalar-field-sensitive mass $m(\varphi)$ ``\`a la" Eardley \cite{Eardley}. 
   The explicit calculation of this black hole ``sensitivity"  includes a constant parameter $\mu$ which identifies to the Schwarzschild mass when the hairs are cut off.\\ 
   
   In this paper,  we will show that this constant $\mu$ can be defined as a function of the entropy of the black hole alone. This can be understood thus: $m(\varphi)$ was chosen not to depend on the gradients of the fields, in particular, on their time derivatives, so that the black hole is moving adiabatically in the fields of its companion; this will imply that it satisfies the first law of thermodynamics;
   moreover, the specific form of $m(\varphi)$ will impose that it exchanges no mass nor charge with its environment.
    Therefore, its entropy will remain constant and hence, can be related to $\mu$.

\section{EMD black holes and their thermodynamics\label{section_thermo}}
The vacuum Einstein-Maxwell-dilaton action of gravity is taken to be, see \cite{Gibbons:1982ih, Gibbons:1987ps,  Garfinkle:1990qj}:
\begin{eqnarray}
16\pi\, I[g_{\mu\nu},A_{\mu},\varphi] & = & \int\! d^{4}x\sqrt{-g}\left(R-2g^{\mu\nu}\partial_{\mu}\varphi\,\partial_{\nu}\varphi-e^{-2a\varphi}F^2\right)\ ,\label{eq:Action1-1}
\end{eqnarray}
where $g$ is the determinant of the metric $g_{\mu\nu}$, $R$ is the Ricci scalar, where $F_{\mu\nu}=\partial_\mu A_\nu-\partial_\nu A_\mu$ with  $F^2=F^{\mu\nu}F_{\mu\nu}$, and where  $a$ parametrizes the theory.

\vfill\eject

The field equations derived from the action (\ref{eq:Action1-1}) are :
\begin{subequations}
\begin{align}
 & R_{\mu\nu}=2\partial_{\mu}\varphi\,\partial_{\nu}\varphi+2e^{-2a\varphi}\left(F_\mu^{\,\,\,\lambda}F_{\nu\lambda}-\frac{1}{4}g_{\mu\nu}F^2\right),\label{eq:eom1}\\
 & D_{\mu}\left(e^{-2a\varphi}F^{\mu\nu}\right)=0\ ,\label{eq:eom3}\\
  & \Box\,\varphi=-\frac{a}{2}e^{-2a\varphi}F^2\ ,\label{eq:eom2}
\end{align}\label{eq:eom}
\end{subequations}
 where $D_\mu$ is the covariant derivative associated to $g_{\mu\nu}$ and $\Box= D_\mu D^\mu$.\\
  
The ``electrically" charged, static, spherically symmetric black hole solutions of the equations above were found in \cite{Gibbons:1982ih, Gibbons:1987ps,  Garfinkle:1990qj}, and read, with $d\Omega^2\equiv d\theta^2+\sin^2\theta\, d\phi^2$,
\begin{align}
&ds^2=-\left(1-\frac{r_+}{r}\right)\left(1-\frac{r_-}{r}\right)^{\frac{1-a^2}{1+a^2}}dt^2+\left(1-\frac{r_+}{r}\right)^{-1}\left(1-\frac{r_-}{r}\right)^{-\frac{1-a^2}{1+a^2}}dr^2+r^2\left(1-\frac{r_-}{r}\right)^{\frac{2a^2}{1+a^2}}d\Omega^2\ ,\label{TNsolution}\\
&A_t=-\sqrt{\frac{r_+ r_-}{1+a^2}}\,\frac{e^{a\varphi_\infty}}{r}\ ,\qquad A_i=0\ ,\qquad\varphi=\varphi_\infty+\frac{a}{1+a^2}\ln\left(1-\frac{r_-}{r}\right)\,,\nonumber
\end{align}
where an irrelevant sign was chosen in the definition of $A_t$. This family of solutions depends on three integration constants : the radius $r_+$ of the horizon, the location $r_-$ of the curvature singularity and the asymptotic value $\varphi_\infty$ of the scalar field which, being associated to no diffeomorphism nor gauge invariance, cannot be set equal to zero and must be identified to the local, adiabiatically varying, value of the scalar field created by the environment, e.g., a faraway companion.\\

The first law of thermodynamics obeyed by these black holes is found in the standard way : 

Their temperature $T$ is defined as
\begin{equation}
T\equiv\frac{\kappa}{2\pi}=\frac{1}{4\pi r_{+}}\left(1-\frac{r_{-}}{r_{+}}\right)^{\frac{1-a^2}{1+a^2}},\label{eq:temperature}
\end{equation}
where $\kappa$ is their surface gravity,
 with $\kappa^{2}=-\frac{1}{2}(\nabla_{\mu}\xi_{\nu}\nabla^{\mu}\xi^{\nu})_{r_{+}}$, and $\xi^\mu=(1,0,0,0)$ being the timelike Killing vector.
 
Their electric potential is
\begin{equation}
\Phi\equiv A_{t}(r\rightarrow \infty)-A_{t}(r_{+})=\sqrt{\frac{r_{-}}{(1+a^2)r_{+}}}e^{a\varphi_{\infty}}\ .\label{eq:potential}
\end{equation}

The action for the metric being Einstein-Hilbert's, the entropy $S$ of the black holes is the fourth of their horizon area $A_+$~: 
\begin{equation}
S\equiv\frac{A_{+}}{4}=\pi r_{+}^{2}\left(1-\frac{r_{-}}{r_{+}}\right)^{\frac{2a^{2}}{1+a^{2}}}.\label{eq:entropy}
\end{equation}

As for the global charges associated to these solutions, that is their electric charge $Q$ and mass $M$, they can be obtained within various approaches, e.g. the Hamiltonian one as developped by Regge-Teitelboim \cite{Regge:1974zd} or the Lagrangian one as developped by Katz \cite{Katz_note_komar,Katz:1996nr}, see appendices \ref{appendixA} and \ref{appendixB} for the explicit computations. They are given by :
\begin{subequations}
\begin{align}
&Q=\sqrt{\frac{r_+ r_-}{1+a^2}}\,e^{-a\varphi_\infty}\ ,\label{globalChargesQ}\\
&M=\frac{1}{2}\left(r_++\frac{1-a^2}{1+a^2}r_-\right)-\frac{a}{1+a^2}\int\! r_- d\varphi_\infty\label{globalChargesM} \ .
\end{align}
\end{subequations}
As usual in scalar-tensor theories of gravity, $M$ is the sum of the ADM mass and of a scalar contribution that is
integrable when given a relationship between $r_-$ and $\varphi_\infty$, see e.g. \cite{Cardenas:2016uzx}, \cite{Anabalon:2016ece} and references therein.

With all these definitions in hand, it is easily checked that the variations of $S$, $Q$, and $M$ with respect to $r_+$, $r_-$ and $\varphi_\infty$, due to exchanges of electric charge or mass between the black hole and its environment, are such that the first law of black hole thermodynamics,
\begin{equation}
T\delta S=\delta M -\Phi \delta Q\,,\label{first_law}
\end{equation}
is satisfied. This first law implies in particular that when the black hole does not exchange any mass ($\delta M=0$) nor charge ($\delta Q=0$) with its environment, its entropy remains constant.

\section{Skeletonized black holes\label{skeletonization}}
In order to address, e.g., the dynamics of a black hole inspiralling around a companion in Einstein-Maxwell-dilaton theories, it was phenomenologically replaced in \cite{Julie:2017rpw} by a point particle described by the following ``skeleton" action:
\begin{equation}
I^{\rm pp}[g_{\mu\nu},A_\mu,\varphi,x^\mu]=I-\int\! m(\varphi)ds+q\int\! A_{\mu}\,dx^{\mu}\,,\label{skelAction}
\end{equation}
where $I$ is given in (\ref{eq:Action1-1}) and where $ds=\sqrt{-g_{\mu\nu} dx^\mu dx^\nu}$, $x^\mu[s]$ being the worldline of the skeletonized black hole. The ``charge parameter" $q$ is taken to be a constant in order to preserve the $U(1)$ symmetry of the full action. As for the scalar-field-sensitive, effective, ``mass" function $m(\varphi)$ (first introduced phenomenologically by Eardley \cite{Eardley}) it is a function of the value of the scalar field at the location of the particle (substracting self divergent terms), created by the faraway companion. The calculation of such mass functions is standard when the compact object is a neutron star, see eg \cite{Damour:1993hw} and \cite{Barausse:2012da}. Let us recall here briefly how it was, for the first time, computed in \cite{Julie:2017rpw} when the compact object is the EMD black hole described in the previous section.\\

The field equations derived from (\ref{skelAction}) are the same as (\ref{eq:eom}) but supplemented by point source terms. They were solved in \cite{Julie:2017rpw} in the rest frame of the particle and at linear order around a background solution consisting of an asymptotically flat spacetime, a vector field which can be ``gauged away" to zero, and an asymptotic scalar field environment $\varphi_\infty$ that is imposed by the faraway companion. The solutions were then identified to the EMD black hole solution (\ref{TNsolution}) at leading, $\mathcal O\left(1/r\right)$, order to yield:
\begin{subequations}
\begin{align}
&q=\sqrt{\frac{r_+r_-}{1+a^2}}\,e^{-a\varphi_\infty}\ ,\label{matching1}\\
&m(\varphi_\infty)=\frac{1}{2}\left(r_++\frac{1-a^2}{1+a^2}\,r_-\right)\ ,\label{matching2}\\
&\frac{dm}{d\varphi}(\varphi_\infty)=\frac{a\, r_-}{1+a^2}\ .\label{matching3}
\end{align}\label{matching}%
\end{subequations}
Note that $m(\varphi_\infty)$ is the ADM mass, which is now adiabatically varying due to the slowly orbiting companion.

The system (\ref{matching}) is integrable : indeed, expressing $r_+$ and $r_-$ in terms of $m$ and $dm/d\varphi$ using (\ref{matching2}) and (\ref{matching3}), and injecting the result into (\ref{matching1}) gives the first order differential equation (omitting the $\infty$ subscript)
\begin{equation}
\left(\frac{dm}{d\varphi}\right)\left(m(\varphi)-\frac{1-a^2}{2a}\frac{dm}{d\varphi}\right)=\frac{a}{2}\,q^2e^{2a\varphi}\ ,\label{eqdiff}
\end{equation}
whose solution depends on the ``charge" parameter $q$, together with another integration constant $\mu$, which can be tentatively called a ``mass" parameter. For example, in the simple case $a=1$, the solution is
\begin{equation}
m(\varphi)=\sqrt{\mu^2+q^2\frac{e^{2\varphi}}{2}}\ .\label{massFunction}
\end{equation}
This is an example, found and studied in \cite{Julie:2017rpw}, of what can be called a hairy black hole ``sensitivity", in keeping with the literature on neutron stars in scalar-tensor theories.

\section{Thermodynamics versus dynamics of EMD black holes \label{ThermovsDynamics}}

Let us now see how the first law of thermodynamics, established in section \ref{section_thermo}, justifies the proposed phenomenological skeletonization of our black hole and provides an interpretation of the constants $q$ and $\mu$ that characterize it.

Comparing  (\ref{globalChargesQ}) and (\ref{matching1}) we first see that we must identify the constant $q$ entering the skeleton action (\ref{skelAction}) to the global electric charge $Q$ of the black hole. The significance of this identification is that the dynamical evolution of the skeletonized black hole is such that its charge remains constant, $\delta Q=0$.

Second, the variation of the global black hole mass $M$, when interacting with its environment, follows from (\ref{globalChargesM}) and reads
\begin{equation}
\delta M=\frac{1}{2}\delta\left(r_++\frac{1-a^2}{1+a^2}r_-\right)-\frac{a\, r_-}{1+a^2}\,\delta\varphi_\infty\ ,
\end{equation}
which is zero when taking into account (\ref{matching2}) and (\ref{matching3}). This means that the black hole mass remains constant as well during its dynamical evolution, $\delta M=0$.

Therefore, the phenomenological skeletonization of black holes proposed in (\ref{skelAction}) amounts to describing them as remaining isolated when, for example, they orbit around a companion.

Finally, the first law (\ref{first_law}) tells us that, since $\delta Q=0$ and $\delta M=0$, the entropy of the black hole remains constant as well ~: $S=const$. Therefore, it must always be possible to define the parameter $\mu$ appearing in the ``mass" function $m(\varphi)$ when integrating (\ref{eqdiff}) as a function of the entropy $S$ only. That this is indeed the case can be seen on the simple example $a=1$, for which $m(\varphi)$ is explicitly given by (\ref{massFunction}), so that, from (\ref{matching2}) : $\mu^2+q^2e^{2\varphi}/2=r_+^2/4$, with $q=Q$. Inverting then (\ref{eq:entropy}) and (\ref{globalChargesQ}) to express $r_+$ in terms of $Q$ and $S$, then yields $\mu^2=S/4\pi$ so that 
\begin{equation}
m(\varphi)=\sqrt{{S\over4\pi}+Q^2{e^{2\varphi}\over2}}\quad (\hbox{case}\ a=1)\,.\label{eq:mu}
\end{equation}
Note that when $r_-=0$ ($\forall a$) the black hole solution is Schwarzschild's so that $m(\varphi)$ is reduced, as it should, to its (constant) mass $m=\sqrt{S/4\pi}=r_+/2$. The  same is true in the Reissner-Nordstr\"om limit, $a=0$, for which $m=(r_++r_-)/2$. On the other hand, when a non-trivial scalar field is present, the phenomenologically, ``Eardley-inspired", scalar-field-sensitive mass function $m(\varphi)$ for black holes which was shown in \cite{Julie:2017rpw} to satisfy (\ref{eqdiff}) is in fact justified by their thermodynamics, and the parameters $q$ and $\mu$ become related to their global electric charge and entropy.

\section{Conclusion}

The results above indicate that the conservative dynamics of a (hairy) black hole when skeletonized ``\`a la" Eardley, as in (\ref{skelAction}), is generically such that it does not exchange energy (nor electric charge) with its environment. Therefore, because of the first law of thermodynamics, the black hole adiabatically readjusts its equilibrium configuration in such a way that its entropy (or area in the case at hand) remains constant. We conjecture that this fact holds in any scalar-vector-tensor theory of gravity and that the scalar-field-sensitive mass functions attributed to skeletonized black holes must guarantee that their Wald entropy \cite{Wald:1999vt} remains constant in their motion around their companion.\\

Of course, our results no longer apply in the late stages of a binary system coalescence. In particular, when the period of the orbit becomes comparable to the readjustment time of a black hole, the adiabatic approximation breaks down and the entropy must increase. Perhaps a way to capture this phenomenon could be to generalize our effective point particle ansatz by introducing a more elaborate one depending for example on the four-gradient of the scalar field as well \cite{Damour:1998jk}. We leave this to further work. Another extension of our work would be to see how spins can be included.\\

Finally, the scalar environment of the black hole could be imposed by a time dependent cosmological environment, rather than a faraway companion. Our results show that in that case as well, the readjustment of the black hole is always such that, at the adiabatic approximation, its entropy remains constant.

\section*{Acknowledgments}
M.C. acknowledges financial support given by Becas Chile, CONICYT.

\vskip 0.5cm
\appendix

\section{Global charges in a Hamiltonian framework\label{appendixA}}

The Hamiltonian generator associated to the Lagrangian \eqref{eq:Action1-1}
reads
\begin{equation}
H\left[\eta^{\perp},\eta^i,\eta^{(A)}\right]=\int d^{3}x\left(\eta^{\perp}\mathcal{H}_{\perp}+\eta^{i}\mathcal{H}_{i}-\eta^{(A)}\mathcal{G}\right)+Q\left[\eta^{\perp},\eta^i,\eta^{(A)}\right]\,.\label{eq:Ham-Gen}
\end{equation}
It is obtained from the standard ADM Hamiltonian (from which the eom can be deduced) by replacing the Lagrange multipliers (that is, $A_t$ together with the lapse $N$ and shift $N^i$ of the standard 1+3 decomposition of the metric) by the asymptotic surface spacetime deformations $\eta^{\perp},\eta^{i}$,
and the gauge parameter of the Abelian symmetry $\eta^{(A)}$.

\vfill\eject

 The constraints $\mathcal{H}_{\perp}, \mathcal{H}_{i}$ and $\mathcal{G}$ are given by :
 \begin{eqnarray}
\mathcal{H}_{\perp} & = & \frac{16\pi}{\sqrt{\gamma}}\left(\pi^{ij}\pi_{ij}-\frac{1}{2}\left(\pi_{\,\,i}^{i}\right)^{2}\right)-\frac{\sqrt{\gamma}}{16\pi}R^{\left(3\right)}
 +2\pi\frac{\pi_{\varphi}^{2}}{\sqrt{\gamma}}+\frac{\sqrt{\gamma}}{8\pi}\partial^{i}\varphi\partial_{i}\varphi-2\pi e^{2a\varphi}\frac{\pi^{i}\pi_{i}}{\sqrt{\gamma}}+\frac{\sqrt{\gamma}}{16\pi}e^{-2a\varphi}F^{ij}F_{ij}\ ,\\
\mathcal{H}_{i} & = & 2\nabla_{j}\pi_{\,\,i}^{j}+\pi_{\varphi}\partial_{i}\varphi+\pi^{j}F_{ij}\ ,\qquad
\mathcal{G}  =  \partial_{i}\pi^{i}\ .
\end{eqnarray}
Here $\gamma_{ij}$ is the spatial metric with determinant $\gamma$, scalar curvature $R^{\left(3\right)}$  and covariant derivative $\nabla_i$, and the conjugate momenta of $\gamma_{ij}$, $\varphi$ and $A_i$ (with $F_{ij}=\partial_i A_j-\partial_jA_i$ and a dot representing time derivatives) are
\begin{align}
&\pi^{ij}=-\frac{\sqrt{\gamma}}{16\pi}\left(K^{ij}-\gamma^{ij}K\right)\quad\hbox{where}\quad K_{ij}=\frac{1}{2N}\left(\nabla_{i}N_{j}+\nabla_{j}N_{i}-\dot{\gamma}_{ij}\right)\,.\\
&\pi_{\varphi}=\frac{\sqrt{\gamma}}{4\pi N}\left(\dot{\varphi}-N^{i}\partial_{i}\varphi\right)\quad,\quad \pi^{i}=-\frac{\sqrt{\gamma}e^{-2a\varphi}}{4\pi N}\left(-\gamma^{ij}F_{0j}+N^{j}\gamma^{ik}F_{jk}\right)\,.
\end{align}
As for the variation $\delta Q$ of the surface term $Q$, it is obtained by demanding that the on-shell variation of the Hamiltonian, with respect to the dynamical variables and $\eta^{\perp},\eta^{i},\eta^{(A)}$, vanishes : $\delta H=0$.
We get $\delta Q=\delta Q^g+\delta Q^\varphi+\delta Q^{(A)}$ with
\begin{align}
&\delta Q^{g}  =  \lim_{r\rightarrow\infty}\frac{1}{16\pi}\intop dS_{l}\,G^{ijkl}\left(\eta^{\perp}\nabla_{k}\delta\gamma_{ij}-\partial_{k}\eta^{\perp}\delta\gamma_{ij}\right)+ \intop dS_{l}\left[2\eta_{k}\delta\pi^{kl}+\left(2\eta^{k}\pi^{jl}-\eta^{l}\pi^{kj}\right)\delta\gamma_{jk}\right],\nonumber\\
&\quad\text{with}\qquad G^{ijkl}=\frac{1}{2}\sqrt{\gamma}\left(\gamma^{ik}\gamma^{jl}+\gamma^{il}\gamma^{jk}-2\gamma^{ij}\gamma^{kl}\right),\\
&\delta Q^{\phi}  = -\lim_{r\rightarrow\infty}\intop dS_{i}\left(\frac{\sqrt{\gamma}}{4\pi}\eta^{\perp}\partial^{i}\varphi\delta\varphi+\eta^{i}\pi_{\varphi}\delta\varphi\right),\nonumber\\
&\delta Q^{(A)}  = - \lim_{r\rightarrow\infty}\intop dS_{i}\left[\frac{\sqrt{\gamma}}{4\pi}\eta^{\perp}e^{-2a\varphi}F^{ij}\delta A_{j}+\left(\eta^{i}\pi^{j}-\pi^{j}\eta^{i}\right)\delta A_{j}-\eta^{(A)}\delta\pi^{i}\right].\nonumber
\end{align}

For the black hole spacetimes (\ref{TNsolution}), the timelike Killing vector is $\xi^\mu=(1,0,0,0)$, so that the only deformation parameters to consider are  $\eta^{(A)}$ and
\begin{eqnarray}
\eta^{\perp} & = & N\xi^{t}\,,
\end{eqnarray}
(where $N$ is $1$ at infinity) and one finally obtains :
\begin{equation}
\delta Q=\xi^{t}\left[\frac{1}{2}\left(\delta r_{+}+\frac{1-a^{2}}{1+a^{2}}\,\delta r_{-}\right)-\frac{a\,r_{-}}{1+a^{2}}\,\delta\varphi_{\infty}\right]-\eta^{(A)}\delta\left(\sqrt{\frac{r_{+}r_{-}}{1+a^{2}}}e^{-a\varphi_{\infty}}\right).\label{eq:deltacarga2}
\end{equation}
In this approach, the variation $\delta M$ of the mass is the coefficient of $\xi^t$, and the variation $\delta Q$ of the electric charge is the coefficient of $-\eta^{(A)}$, see \cite{Cardenas:2016uzx}. Hence, integration yields
\begin{align}
M & =\frac{1}{2}\left(r_{+}+\frac{1-a^{2}}{1+a^{2}}\, r_{-}\right)-\frac{a\,}{1+a^{2}}\,\int r_{-}\delta\varphi_{\infty}\ ,\label{eq:dM}\\
Q & =\sqrt{\frac{r_{+}r_{-}}{1+a^{2}}}e^{-a\varphi_{\infty}}\ .\label{eq:dQ}
\end{align}

\section{Mass as a Noether charge\label{appendixB}}

The Einstein-Maxwell-dilaton-Katz action is defined by
\begin{equation}
16\pi\,I_{\rm K}=\int\! d^{4}x\sqrt{-g}\left(R-2g^{\mu\nu}\partial_{\mu}\varphi\,\partial_{\nu}\varphi-e^{-2a\varphi}F^2\right)-\int\! d^4x\sqrt{-\bar g}\bar R+\int\! d^{4}x\,\partial_{\mu}\left(\hat k_{K}^{\mu}+\hat k_{S}^{\mu}\right)\ ,\label{EMDK_action}
\end{equation}
where barred quantities refer to a reference spacetime that will be taken to be flat, see \cite{Anabalon:2016ece}. The last term is (a hat meaning multiplication by $\sqrt{-g}$)
\begin{align}
\hat k_{K}^{\mu}=-\left(\hat g^{\nu\rho}\Delta_{\nu\rho}^{\mu}-\hat g^{\mu\nu}\Delta_{\nu\rho}^{\rho}\right)\quad\text{with}\quad\Delta_{\nu\rho}^{\mu}=\Gamma_{\nu\rho}^{\mu}-\bar{\Gamma}_{\nu\rho}^{\mu}\qquad\text{and}\qquad \hat k_{S}^{\mu}=A\,\hat\partial^{\mu}\varphi\ ,
\end{align}
where $\Gamma_{\nu\rho}^{\mu}$ and $\bar{\Gamma}_{\nu\rho}^{\mu}$ are the Christoffel symbols associated to $g_{\mu\nu}$ and $\bar g_{\mu\nu}$ and where $A$ is an arbitrary constant.\\

 The variation of (\ref{EMDK_action}) with respect to the scalar field reads, on-shell:
\begin{align}
\label{EMDK_variation}
 16\pi\,\delta_{\varphi}I_{\rm K}=\int\! d^{4}x\,\partial_{\mu}\hat{V}_{\varphi}^{\mu}
 \qquad\text{with}\qquad\hat{V}_{\varphi}^{\mu}=-4(\hat{\partial}^{\mu}\varphi)\delta\varphi+A\,\hat g^{\mu\nu}\delta(\partial_{\nu}\varphi)\ .
\end{align}
For the black hole family (\ref{TNsolution}) we have, upon variation of the integration constants $r_+$, $r_-$ and $\varphi_\infty$,
\begin{align}
 16\pi\,\delta_\varphi I_K^{\rm onshell}=\lim_{r\rightarrow\infty}\int_{t_1}^{t_2}\! dt\int_{S^2}\!d\theta\,d\phi\, \hat V_{\varphi}^{r}\qquad\text{with}\qquad \hat V_{\varphi}^{r}=\frac{a\,\sin\theta}{1+a^2}\left(A\,\delta r_{-}-4\,r_-\,\delta\varphi_{\infty}\right)+\mathcal{O}\left(1/r\right)\ .
\end{align}
Imposing now that the variation of the action vanishes for the broadest
possible family of black holes yields a relationship between $r_{-}$ and $\varphi_{\infty}$:
\begin{equation}
A\,\delta r_{-}=4\,r_-\,\delta\varphi_{\infty}\ .\label{restriction}
\end{equation}
Note that no further conditions emerge from varying (\ref{EMDK_action}) with respect to $g_{\mu\nu}$ and $A_\mu$; indeed one finds $\hat V^r_g=\mathcal O(1/r)$ and $\hat V^r_A=\mathcal O(1/r)$ on shell, and at spatial infinity.\\

Finally, exploiting the invariance of (\ref{EMDK_action}) under diffeomorphisms, $x^\mu\rightarrow x^\mu+\xi^\mu$, one obtains ``\`a la Noether" a conserved current $\hat j^\mu$ deriving from the generalized Katz-Bicak-Lynden-Bell (KBL) superpotential, see, e.g. \cite{Deruelle:2003ps}:
\begin{equation}
\partial_\mu\hat j^\mu=0\qquad\text{with}\qquad\hat j^\mu=\partial_\nu\hat J^{[\mu\nu]} \qquad\text{and}\qquad 8\pi\hat{J}^{[\mu\nu]}=\nabla^{[\mu}\hat{\xi}^{\nu]}-\overline{\nabla^{[\mu}\hat{\xi}^{\nu]}}+\xi^{[\mu}\hat{k}_{K}^{\nu]}+\xi^{[\mu}\hat{k}_{S}^{\nu]}\ .\label{KBL_superpotential}
\end{equation}
When spacetime is stationary, $\xi^\mu\equiv(1,0,0,0)$ is the timelike Killing vector and the mass of the black hole (\ref{TNsolution}) is defined as 
\begin{align}
&M=-\lim_{r\rightarrow\infty}\int\! d\theta d\phi\,\hat{J}^{[0r]}=M_K+M_S\\
\text{where}\quad &M_{K}=\frac{1}{2}\left(r_++\frac{1-a^2}{1+a^2}r_-\right)\quad\text{and}\quad M_S=-\frac{A}{4}\frac{a\,r_-}{1+a^2}\ .\nonumber
\end{align}
Note that the Katz mass $M_K$ coincides with the ADM mass. Finally, $M$ is easily written independently of the constant $A$, using (\ref{restriction}) as:
\begin{equation}
M=\frac{1}{2}\left(r_++\frac{1-a^2}{1+a^2}r_-\right)-\frac{a}{1+a^2}\int r_- d\varphi_\infty\ ,
\end{equation}
which coincides with (\ref{eq:dM}).

\vskip 1cm

\bibliographystyle{unsrt}

\end{document}